# A privacy-preserving data storage and service framework based on deep learning and blockchain for construction workers' wearable IoT sensors

Xiaoshan Zhou [1], Ankit Kumar Shaw [2], and Pin-Chao Liao [1]

[1] Department of Construction Management, Tsinghua University, Beijing, 10084 China
[2] School of Vehicle and Mobility, Tsinghua University, Beijing, 10084 China

Corresponding author: Pin-Chao Liao (e-mail: pinchao@mails.tsinghua.edu.cn).

This research was funded by National Natural Science Foundation of China (No., 51878382).

**ABSTRACT** Classifying brain signals collected by wearable Internet of Things (IoT) sensors, especially brain-computer interfaces (BCIs), is one of the fastest-growing areas of research. However, research has mostly ignored the secure storage and privacy protection issues of collected personal neurophysiological data. Therefore, in this article, we try to bridge this gap and propose a secure privacy-preserving protocol for implementing BCI applications. We first transformed brain signals into images and used generative adversarial network to generate synthetic signals to protect data privacy. Subsequently, we applied the paradigm of transfer learning for signal classification. The proposed method was evaluated by a case study and results indicate that real electroencephalogram data augmented with artificially generated samples provide superior classification performance. In addition, we proposed a blockchain-based scheme and developed a prototype on Ethereum, which aims to make storing, querying and sharing personal neurophysiological data and analysis reports secure and privacy-aware. The rights of three main transaction bodies – construction workers, BCI service providers and project managers – are described and the advantages of the proposed system are discussed. We believe this paper provides a well-rounded solution to safeguard private data against cyber-attacks, level the playing field for BCI application developers, and to the end improve professional well-being in the industry.

## I. INTRODUCTION

Brain-computer interface (BCI) applications for construction workers have grown dramatically in recent years. These systems continuously monitor the users' electroencephalographic (EEG) data and interpret the brain signals to detect fatigue [1], vigilance and attention levels [2], mental workload [3], stress [4], and emotions [5], etc. Early warning signs related to their safety issues are provided to the users detected under a subnormal state to mitigate health risks and safety hazards on construction sites [6]. Although state-of-art performance, such as plug-and-play control of a BCI [7], has been achieved, large quantities of personal data are required for developing robust machine learning (ML) algorithms in applications involving the BCI. However, it is time-consuming and inconvenient to collect high-quality EEG data due to several reasons, such as long calibration time [8], between-subject variability [9], and data corruption as a result of various experimental factors [10], especially for portable BCIs with limited computational resources [11].

Thus, if personal data can be shared in a secure manner, it will benefit all related stakeholders, including the wearable device users (i.e., the construction workers), project managers, researchers, and companies. As personal assets, the data collected by sensors should be controlled by the users themselves. However, they are primarily controlled by third parties such as service providers and device manufacturers or dispersed across multiple information management systems [12], which makes data sharing harder and poses privacy and security risks as the vast amount of data collected by the sensors is currently stored in a centralized manner, which is a prime target for cyber-attacks that seek to steal or alter sensitive personal information [10]. These risks have sparked a field called neurosecurity, which is concerned with preventing neural devices from attempting to hack or capture the brain signals in order to ensure the confidentiality, integrity, and availability of data [13].

In this paper, we propose a full-chain private data and analysis report sharing protocol combining blockchain, cloud storage, and deep learning techniques, so that the users can safely share their neurophysiological data in order to get individualized services and to assist researchers and other commercial data users with obtaining training data efficiently and transparently. The main contributions of this paper include: 1) combining blockchain and off-blockchain cloud storage to construct a personal neurophysiological data management platform that focuses on secure storage and privacy; 2) Safeguarding privacy against data leakage by utilizing a generative adversarial network (GAN) to generate synthetic brain signals so as to conceal sensitive characteristics of the original data and improve classification performance; 3) introducing a service framework and developing the corresponding Ethereum blockchain prototype for the scenario of sharing data analysis reports among three types of transaction bodies; 4) highlighting the characteristics of the proposed protocol including user ownership, tamper-proof, convenient interoperability and excellent business opportunity. Henceforth, the rest of the paper is structured as follows: the background and previous studies are described in Section II; the storage schema of neurophysiological data and a service framework based on blockchain technology is described in Section III, whereas Section IV describes in detail the technical implementation of the data augmentation protocol and a case study; Section V contains a prototype of the blockchain-enabled platform and advantage analysis, and concluding remarks and future directions to further this work are found in section VI.

## II. RELATED WORK

Although various wearable BCI applications have been developed to improve health and safety for construction workers, if promoted in a market-oriented way, they bring users, i.e., the construction workers, numerous privacy concerns. As workers do not have ownership of their data collected by BCI applications, they are usually unaware of how the data will be analyzed later. Moreover, current neurophysiological data repositories are often centralized, putting them at risk of various cyber-attacks, such as malicious tampering and hacking [10]. As a result, there is a high risk of personal data loss or hacking, which raises concerns related to data security and personal data privacy. In addition, individuals' data analysis reports provided by BCI service providers have not been managed in an integrated way, making it difficult to track the health status of workers in the long term.

From the perspective of BCI service developers, due to the difficulty of neurophysiological data collection, training models on a small data size makes it challenging to achieve the expected performance for brain signal classification[14]. As the demand for developing wearable devices increases, numerous BCI service providers want to acquire personal neurophysiological data through data exchange and sharing. A possible solution is to encourage users to share data to the public research field; however, limited studies have proposed comprehensive protocols for sharing data in a safe and transparent way to guarantee the integrity and privacy of personal data. Thus, a decentralized, verifiable, and immutable scheme is needed for managing neurophysiological data.

In recent years, blockchain technology has gained substantial popularity in diverse academic fields due to its unique advantages, such as decentralized control, high degree of anonymity, transparency, and immutability [15]. The blockchain could reduce the risk of data breaches by encrypting data, where asymmetric cryptography to authenticate communication between system participants [16]. The blockchain enables multiple parties to share data in real-time. Every operation is verified and accountable in blockchain [17]. Time-stamped events or transactions are recorded in blocks of a long chain where information is tamper-proof permanently. On a nonpermissive blockchain, all records are visible to all parties, whereas a permissive blockchain protects privacy by hiding all identities of parties and using access control to authenticate query permissions. These attributes make blockchain a unique solution for data management concerned with both user privacy and data sharing.

In 2015, the blockchain was introduced as an automated access-control manager that ensures users have ownership and control over their personal data, no longer requiring trust in third parties [18]. Since 2016, blockchain technology has been used to manage data related to healthcare and has become the primary research focus for many researchers [12]. Existing research related to blockchain in healthcare primarily includes health information protection [19], medical data storage and sharing [20], automated remote patient monitoring [21], healthcare predictive modeling [22], etc. For example, the research presented in [23] proposed a blockchain-based application framework named Healthcare Data Gateway, in which the architecture not only allowed patients to retain, manage, and share their information safely, but also included secure multifactor computing to allow external parties to access medical and health information without invading the privacy of the patients. The work in [24] designed a decentralized medical file management system based on blockchain. The patients were given a detailed, unchangeable log and easy access to their Electronic Medical Record from several medical institutions. Another research provided various applications of blockchain innovation to the human services framework, such as accessing and managing large medical databases, using smart contracts for securing the data exchange between the patient and doctor while eliminating the third party [25]. For personal data collected by wearable devices, the research presented in [26] introduced blockchain to ensure a secure data ingress mechanism for physiological data collected by various sensors that are previously aggregated and stored in a personal server. In addition, Sujin et al. proposed a blockchain network to validate the authenticity of electroencephalography (EEG) data using smart contracts [27].

As summarized in Table 1, blockchain technology has been utilized to manage sensitive personal data; however, their schemes are still insufficient for personal neurophysiological data of BCI applications. First, the data type of the previous studies was mainly medical records, which requires less storage space than continuously collected neurophysiological data. Next, the previous studies primarily focused on an algorithmic analysis without specific use cases. Furthermore, most blockchain-based solutions are scheme designs, without implementation details and in-depth discussion on permissions of main transaction bodies. In this article, we focus on the storage of data and sharing needs among multiple stakeholders, integrating both cloud storage and smart contract to a blockchain-based protocol to meet the need of large-scale data exchange, and promotes the active participation of relevant parties in a privacy-preserving manner.

TABLE I
COMPARATIVE ANALYSIS OF RECENT STUDIES OVER DATA MANAGEMENT USING BLOCKCHAIN TECHNOLOGY

| Author | Year | Data Type | Major Findings | Major Shortcomings and Challenges |
|---|---|---|---|---|
| Khatoon [25] | 2020 | Patient waiting list | This is an in-depth review of blockchain-based healthcare systems. In this work, multiple workflows are covered. The associated cost has been estimated by a feasibility study. | This work can be extended to blockchain design and smart contract-based use cases for personal data management systems. System design and implementation details are necessary in near future. |
| Zyskind et al.[18] | 2015 | Not limited | This is the first work that describes a decentralized personal data management protocol using blockchain technology. Moreover, future extensions to blockchains are discussed for trusted computing problems. | Very generic discussions are developed to use blockchain technology for personal data protection. The rapidly expanding domain is expecting in-depth discussions. |
| Sun et al. [28] | 2020 | Electronic medical records | This work has proposed a distributed electronic medical record sharing scheme with security and privacy preservation leveraging blockchain and smart contract technology. | The addressed problems of data storage and sharing can be extended to applications like BCI systems. Likewise, the proposed approach can be tested for various other applications. |
| Chen et al.[29] | 2018 | Medical records | Design a storage scheme to manage personal medical data based on blockchain and cloud storage. A service framework for sharing medical records is also described. | No blockchain and smart contract integrated framework is proposed, or no such use case is delivered to smartly guarantee the trust among system participants. |
| Li et al.[30] | 2020 | Sensing image | This work has proposed a blockchain-based security transmission and storage solution for sensing image in the Internet of Things. Here, an algorithm model is proposed and security analysis is done. | This work is more of an algorithmic analysis rather than framework-based implementation. |
| Bak et al.[27] | 2019 | EEG data | The current work proposes a blockchain platform to fix the personal information vulnerability when a service is attacked. The proposed system is validated by a proof-of-concept. | Integration of multiple advanced technologies and related studies are not discussed in this work. For example, the need of cloud computing, data storage format and cryptography features are not explored. These aspects will be included in the future. |

## III. AN OVERVIEW OF THE PROPOSED PROTOCOL

In order to achieve safe data storage and sharing, an ideal system should bring all stakeholders to a common platform and enable the transparent exchange of data and records between all users, service providers, and other third parties. The blockchain is the platform that manages identity security, transaction transparency, data integrity, and accessibility to all information on the network. The architecture of the blockchain-based system is shown in Figure 1.

### A. DATA COLLECTION

To collect brain signals as input for real-world BCI applications, various non-invasive neuroimaging techniques — such as EEG, near-infrared spectroscopy (NIRS), magnetoencephalography (MEG), and functional magnetic resonance imaging (fMRI) — have been utilized. When choosing the most appropriate technique for brain signal recording, several factors need to be considered, including the temporal-spatial resolution, portability, and commercial viability. Among them, EEG has been most widely used particularly due to its high temporal resolution. NIRS is also promising because it is user-friendly and insensitive to electrical interference from both internal sources and the external environment [31]. In contrast, both MEG and fMRI have relatively high costs and poor portability, thus being mainly limited to the medical field. In the remainder of this article, we selected EEG as a representative modality for collecting brain signals.

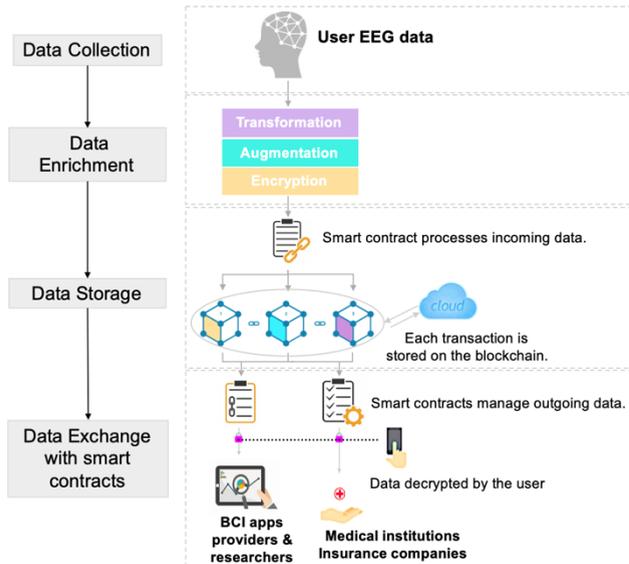

**FIGURE 1.** The architecture of the proposed blockchain-based system.

### B. DATA ENRICHMENT
#### 1) DATA TRANSFORMATION
Once brain signals are collected, the signal is then transformed into an image format for the following classification and online storage. There are numerous papers in the literature that attempt to classify brain signals using ML algorithms. As deep learning with convolutional neural networks (CNNs) has achieved outstanding performance in image classification recently, there is increasing research interest to use CNNs for end-to-end EEG analysis [32]; and see [33] for a comprehensive review. The advantage of transforming EEG signals into images is as follows: so far, most of the research has focused on lab-based circumstances, so the accuracy achieved by these studies is questionable for practical applications. As deep learning has shown enormous potential in the computer vision field, advanced models, though complicated and inexplicable to some extent, provide state-of-art performance for image classification. In this case, we may use the transfer learning approach to take advantage of some of the most prominent classification models in the literature. The rationale here has been backed by sufficient findings that improved performance is witnessed [34, 35].

#### 2) DATA AUGMENTATION
To overcome the difficulties of small training data size, generative adversarial network (GAN) [36] has been applied to generate more artificial EEG signals for data augmentation recently, which has been proven to indeed improve the performance of deep learning models [14, 37]. As such, it can further be used to reduce calibration time to facilitate real-world BCI adoption and restore of corrupted data segments [38]. The generated synthetic EEG brain signals avoid disclosing sensitive characteristics of the original data, fulfilling the privacy protection purpose simultaneously. In healthcare systems, GAN has been used to generate realistic synthetic data for individual participants to mitigate privacy concerns when sharing electronic health record data [39, 40].

#### 3) DATA ENCRYPTION
Augmented brain signals will be encrypted by asymmetric cryptography. A pair of keys, namely a public key and a private key, will be created in this case for encrypting and decrypting, respectively. The signal data encryption is done in ciphertext using the user's public key and then safely saved in the blockchain. The hash function is used to compress any size of input data into a fixed-length output value. Since the hash function never produces repeating values and it is hard to guess the input data based on the output value, it can be regarded as a fingerprint of the input information. As a result, it is easy to determine whether the data has been tampered with or not by comparing the hash. To access the data, one must decrypt it with the associated private key only known to the user.

### C. DATA STORAGE
The primary purpose of introducing blockchain is to guarantee the integrity and security of sensitive EEG data. Most studies that develop blockchain-based healthcare systems to manage personal data focus on relatively static data, such as blood type, history of diseases, etc. Because this type of data requires less storage space, it can be saved and shared directly within the blockchain. However, EEG data is continuously recorded and stored on a millisecond time scale. Such large volumes of dynamic data are difficult to be manipulated within the blockchain. A study [16] provided an approach to integrating cloud storage into blockchain-enabled personal data systems, shedding light on an off-chain storage solution for large neurophysiological data in our scheme. As a result, only EEG data index information, analysis results, and transaction histories are saved in the blockchain, wheras large EEG data are stored in cloud storage that is secured by the blockchain.

### D. DATA EXCHANGE WITH SMART CONTRACTS
As per Figure 2, construction workers, BCI service providers, and project managers are the three main transaction bodies in the blockchain. Construction workers are users of BCI applications who have ownership and access to their neurophysiological data. BCI service providers are responsible for the diagnosing abnormal healthy states (both physically and mentally) of users and for generating their analysis reports. Project managers enjoy the decision-making power of selecting BCI app services and can query analysis reports of their subordinate workers. Some third-party agencies (insurance companies, medical institutions, etc.) may also get involved in the blockchain system. For example, workers' analysis reports can be obtained through a blockchain-based review and trusting system; in case of an accident, the BCI service provider will automatically send information to emergency services.

Access control, data query, and report update are the main transactions in the blockchain. Smart contract, a protocol in digital format, is introduced to the blockchain to ensure good trust and commitment fulfillment between trading partners [41]. The construction worker, for example, has complete control over the usage of personal neurophysiological data. Access control through smart contracts allows the workers to grant other parties permission for accessing their personal data. The accessibility agreement that empowers data exchange is reached without the participation of intermediaries.

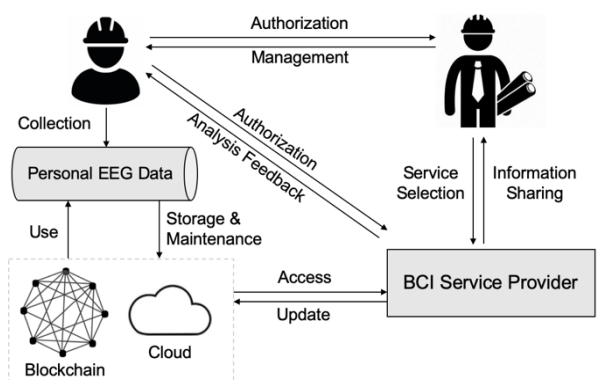

**FIGURE 2.** The service framework among three main types of transaction bodies.

## IV. DATA AUGMENTATION IMPLEMENTATION: A CASE STUDY

### A. EEG DATA COLLECTION

We designed a hazard recognition task in which participants viewed construction scenario pictures present on the screen and judged whether a hazard existed by pressing corresponding keys. Figure 3 shows the experimental protocol. The experiment's details can be found in [42]. The EEG data were recorded from seventy-seven construction workers. For each participant, two recording sessions were undertaken (the official session and the validation session). The official session was divided into four blocks, each of which contained 30 trials and was separated by short breaks. There were 120 trials, including 60 pairs of construction scenarios, each with two opposite conditions (hazardous or safe). The validation session was composed of 30 trials randomly chosen from the preceding 120 trials. The consistency of responses to the same stimuli was examined, and subjects were removed if the inconsistency rate was greater than 50%.

Thirty-two electrodes (Neuroscan system in China and Brain Products in the United States, according to the international 10–20 system) were used to continuously record EEG signals at a sampling rate of 250 Hz with all electrodes referenced. The impedances of the electrodes were kept below 20 kΩ.

The signals were then bandpass filtered between 0.1 and 40 Hz, and independent component analyses were used to remove any artifacts (eye movements, muscle noise, line noise, etc.). Each participant's trial was then segmented into a 1200-ms epoch with a 200-ms pre-stimulus baseline.

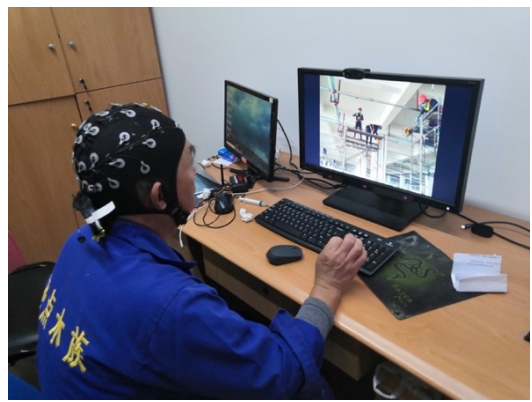

**FIGURE 3.** The experiment of a hazard recognition task with EEG data collected.

### B. BRAIN SIGNAL PROCESSING DETAILS

In this section, the technique for converting brain signals to images is discussed. Then synthetic EEG data were generated using GAN. Finally, we looked into the paradigm of transfer learning to classify EEG signals. See Figure4 for a flowchart of the study methodology. Some implementation details are described as follows.

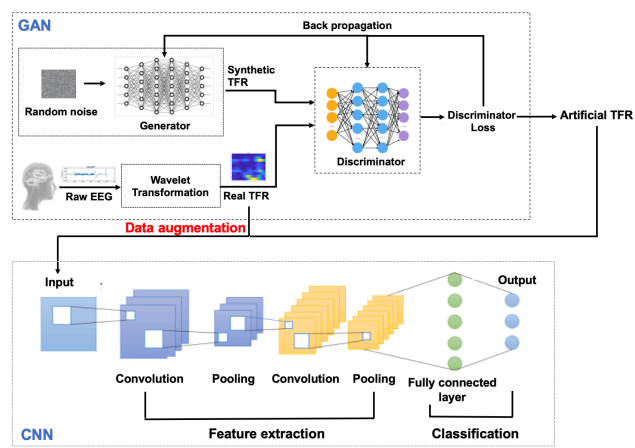

**FIGURE 4.** The analysis flowchart.

#### 1) CONVERTING BRAIN SIGNALS TO IMAGES

We used time-frequency analyses based on a wavelet transform approach [42] to convert EEG signals into images. We performed the analysis using the FieldTrip toolbox [43] in MATLAB (MathWorks, Inc., Natlick, MA, USA) and used a Hanning taper with a 500-ms sliding window in 50-ms time steps to compromise between time and frequency resolutions. The time-frequency representations (TFRs) of EEG signals were computed for each trial of each participant.

#### 2) GENERATING SYNTHETIC BRAIN SIGNAL SAMPLES

GAN was applied for artificial EEG data generation. GAN is made up of two neural networks: the Generator (G), which is in charge of creating artificial (synthetic) samples, and the Discriminator (D), which is in charge of discriminating whether the sample is real or artificial. G's purpose is to trick the discriminator to the point where the discriminator can not tell the difference between the real and artificial samples. Equation (1) can be used to formulate this adversarial optimization issue.

$$min_G \, max_D(D, G) = E_{x \sim p_{(x)}}[\log D(x)] + E_{z \sim p_{(z)}}[\log(1 - \log D(G(z)))] \quad (1)$$

such that G and D are the Generator parameters and the Discriminator parameters, respectively. x is real samples. G network generates the artificial sample from a stochastic noise input z. $D(x)$ is the probability of $x$ that belongs to the real or the artificial data distributions.

Tables 2 and 3 summarize the architecture of the proposed privacy-preserving GAN approach for EEG data augmentation. The proposed GAN's G network is composed of four main blocks, each of which has one up-sampling and one convolution layer with a ReLU activation function and an Adam optimizer for activation and optimization, respectively. For better classification performance, we used a CNN for the D network that learns spatio-temporal patterns in the raw EEG data. The CNN is composed of six processing blocks. The first five blocks are primarily used to learn spatio-temporal features from EEG data, whereas the last block is a standard fully-connected network that is utilized for final classification. Figure 5 shows the generator and the discriminator losses over 500 epochs for a subject. Examples of artificial time-frequency map samples generated over 50, 100, 200, and 500 epochs, respectively are shown in Figure 6.

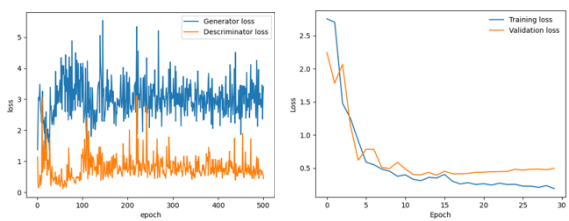

**FIGURE 5.** GAN loss (left) and CNN classification loss (right).

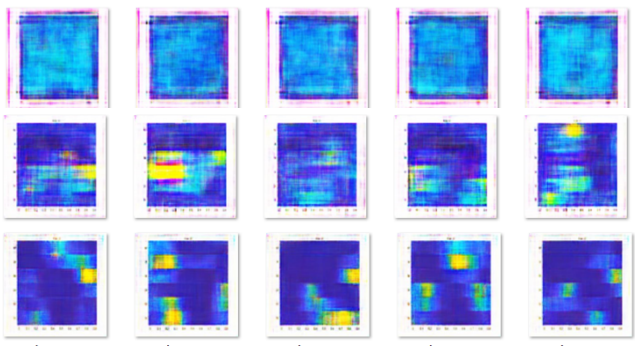

**FIGURE 6.** Examples of artificial EEG samples over 50, 100, 200, 500 training epochs respectively (from top to bottom).

TABLE II

| GENERATOR ARCHITECTURE | |
|---|---|
| Layer | Output shape |
| Dense | 4096 |
| Reshape | (4, 4, 256) |
| Upsampling2D | (8, 8, 256) |
| Conv2D | (8, 8, 256) |
| Batch Normalization | (8, 8, 256) |
| Activation-ReLU | (8, 8, 256) |
| Upsampling2D | (16, 16, 256) |
| Conv2D | (16, 16, 256) |
| Batch Normalization | (16, 16, 256) |
| Activation- ReLU | (16, 16, 256) |
| Upsampling2D | (32, 32, 128) |
| Conv2D | (32, 32, 128) |
| Batch Normalization | (32, 32, 128) |
| Activation- ReLU | (32, 32, 128) |
| Upsampling2D | (96, 96, 128) |
| Conv2D | (96, 96, 128) |
| Batch Normalization | (96, 96, 128) |
| Activation- ReLU | (96, 96, 128) |
| Conv2D | (96, 96, 3) |
| Activation-Tanh | (96, 96, 3) |

TABLE III

| DISCRIMINATOR ARCHITECTURE | |
|---|---|
| Layer | Output shape |
| Conv2D | (48, 48, 32) |
| LeakyReLU | (48, 48, 32) |
| Dropout | (48, 48, 32) |
| Conv2D | (24, 24, 64) |
| ZeroPadding2D | (25, 25, 64) |
| Batch Normalization | (25, 25, 64) |
| Dropout | (25, 25, 64) |
| Conv2D | (13, 13, 128) |
| Batch Normalization | (13, 13, 128) |
| LeakyReLU | (13, 13, 128) |
| Dropout | (13, 13, 128) |
| Conv2D | (13, 13, 256) |
| Batch Normalization | (13, 13, 256) |
| LeakyReLU | (13, 13, 256) |
| Dropout | (13, 13, 256) |
| Conv2D | (13, 13, 512) |
| Batch Normalization | (13, 13, 512) |
| LeakyReLU | (13, 13, 512) |
| Dropout | (13, 13, 512) |
| Flatten | 86528 |
| Dense-Sigmoid | 1 |

### 3) DATA CLASSIFICATION

We trained classifiers to distinguish brain activities induced by hazardous stimuli from those induced by safe stimuli from four subjects. According to [44], there is a strong link between task accuracy performance and the discriminability of brain activation induced by two types of stimuli. Therefore, we included four subjects who demonstrated task accuracy performance in quartiles (Subject One: 73.33%; Subject Two: 65%; Subject Three: 60.83%; Subject Four: 45.83%) in the following analysis. Also, refer to [44], the optimal spectral and spatial EEG characteristics for task-related performance classification are gamma-band (30-40 Hz) and PO8 electrode.

*a) Feature extraction, selection and classification*

As we have transformed EEG epochs into the time-frequency domain, we used TFRs (denoted as "power" afterwards as suggested by [45]) of gamma band from PO8 electrode in a 100ms time sliding window for features. This is equivalent to extracting the value of each pixel in the time-frequency map and averaging over the corresponding time-frequency window to get the temporal-spectral features of EEG data. To assess the performance of classification, we split the real data into train/test subsets according to a split ratio of 70%:30%. Since the extracted features are high-dimensional with redundant information, Principal Component Analysis, which is a representative dimensionality reduction algorithm, was used to avoid the overfitting of the training data. The top three components were chosen for the subsequent analysis. Then we utilized various ML algorithms, including Support Vector Machine (SVM), Logistic Regression (LR), K-Nearest Neighbor (KNN), Decision Tree (DT), Gaussian

Naïve Bayes (NB), and Multilayer Perceptron (MLP) for classification.

*b) CNN-based transfer learning*

We can recall from Section 4.1, where we transformed EEG signals into images. In this section, we will conduct signal classification utilizing the transfer learning paradigm for image classification. The idea behind transfer learning is that we train a model on a dataset (say $D_s$) for a classification problem (say $C_a$). Now, instead of having to train the model from scratch for a different classification problem (say $C_b$), we can use the model that was trained on $D_s$ and apply the learned knowledge to the problem $C_b$. Thus, we make use of existing knowledge. In this article, we applied ResNet50, one of the most used CNN models for classification. The model, which was trained on 1.2 million images from the ImageNet dataset [46], has performed well in a number of challenging and difficult situations. As a result, the ImageNet dataset became the model's source domain, while the brain signals that had been converted into images became the model's target domain.

The ResNet50 was trained on two types of data: real EEG data and real data augmented with synthetic samples. To examine the effect of the ratio of artificial to real data on classification results and the effect of the GAN training epoch, we constructed five augmented training data sets for each GAN training epoch, with the ratio of synthetic to real data 0.2, 0.5, 0.8, 1, 1.5, respectively. In addition to the original dataset, we trained the ResNet50 on each of the twenty augmented datasets and reported the classification accuracy on the corresponding test set. To avoid the overfitting of training data, we chose 20 epochs to train the CNN for classification according to Figure 5.

4) RESULTS

Table 4 shows the classification accuracy for models that were trained on real data only. It is observable that the proposed method, i.e., the paradigm of CNN-based transfer learning, performed better than representative ML algorithms in terms of classification accuracy. The subject with the highest task accuracy performance also provided the highest brain signal classification accuracy, supporting our hypothesis.

TABLE IV
CLASSIFICATION ACCURACY FOR MODELS

|  | CNN | SVM | LR | DT | NB | MLP | ML average |
|---|---|---|---|---|---|---|---|
| Sub1 | 58.3 | 47.2 | 50.0 | 61.1 | 50.0 | 54.2 | 52.50 |
| Sub2 | 54.2 | 52.8 | 50.0 | 47.2 | 47.2 | 50.5 | 49.54 |
| Sub3 | 54.2 | 50.6 | 52.8 | 44.4 | 52.8 | 50.9 | 50.30 |
| Sub4 | 50.0 | 47.2 | 38.9 | 55.6 | 47.2 | 48.2 | 47.42 |

We followingly selected Subject One to evaluate the impact of data augmentation implemented by GAN. Figure 7 shows the classification accuracy for the ResNet50 model trained on real data only and on a combination of real data and artificially generated data. As shown in Figure 7, training data augmented with appropriate ratios provided improved classification accuracy. Specifically, the results showed that adding artificial samples from 50 epochs of GAN training significantly increased the accuracy from a 58.33% baseline (trained with the real data only) to 86.67% (trained with data augmented with a 1.5 ratio of artificial to real data). Although the quality of the samples generated by GAN using a small number of epochs is not optimal, it gave superior classification performance. Conversely, the addition of a low ratio of relatively high-quality artificial samples deteriorated the performance probably because it produced noise in the models. Increasing the ratio of artificial to real data can compensate the problem; however, it is not time- and cost-saving.

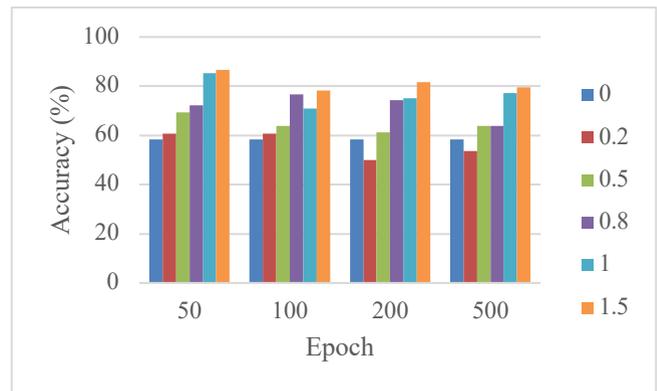

FIGURE 7. The impact of the ratio of artificial data to real data and training epoch of GAN on classification accuracy.

## V. Prototype system and advantages: a proof-of-concept

In this study, we developed a private blockchain project on Ethereum with Truffle and Ganache GUI. The Ethereum blockchain is a decentralized and open-source platform that allows creation smart contract-based projects. Smart contract is a legal agreement among various stakeholders to maintain the credibility and integrity of the system. It is typically written using higher lever programming language like Solidity. Smart contract runs on the Ethereum Virtual Machine (EVM), a decentralized network made up of Ethereum nodes responsible for running EVM instances. These contracts can self-execute when required criteria are met. Here the mechanism behind it is as follows: the smart contract will be first compiled into bytecode, with each byte representing an operation, then recorded in the blockchain as an EVM transaction and subsequently picked up by a block. Therefore, any operations initiated by authorized identities will be recorded in the blockchain, where status is refreshed to new blocks. In our prototype demo, the front-end is implemented with Flask, and Web3.py is used to connect the back-end and the front-end.

In the blockchain, the main functions include registration, access control over private EEG data, and the release and view of the individual analysis report.

1) Registration: All related parties can register identities in the system. When their personal information is filled in on the registration fleet, the system will generate a blockchain address and a contract address (for construction workers only) and send them to the issuer, which is simultaneously recorded in the blockchain (as shown in Figure 8).

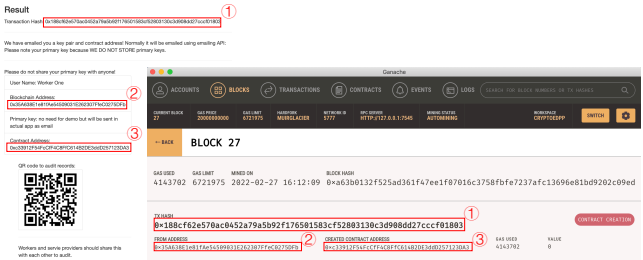

**FIGURE 8.** The registration feedback for construction workers, which is recorded on blockchain.

2) Access control: The usage rights of the neurophysiological data and the view permissions of the analysis report are entirely controlled by the construction workers. The workers can authorize other parties, such as BCI service providers' access to their private EEG data by inputting the account address of the identity and clicking on the "Grant" button, causing the result view to appear shown in Figure 9. They can also withdraw their authorization at any time. Hence, if an obscure identity without authorization queries the private data, the system will deny the request automatically. In practice, we propose that workers make a service appointment with BCI service providers on a daily basis, which will result in a unique report ID (as shown in Figure 1). This report ID will later be used by BCI service providers to update the analysis report or by project managers to view the analysis report of their subordinate workers.

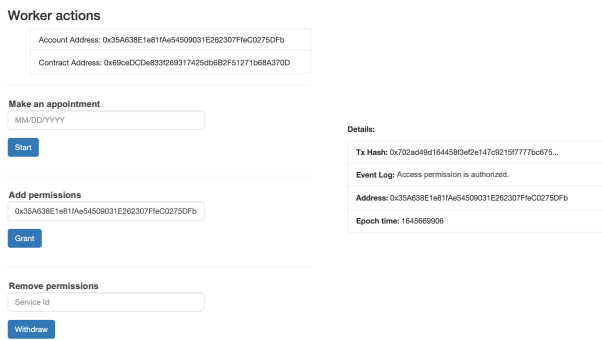

**FIGURE 9.** Grant access permission

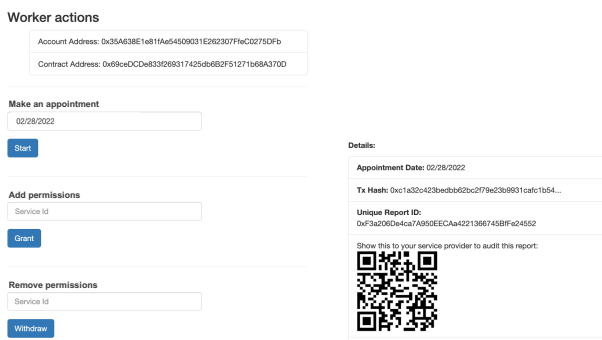

**FIGURE 10.** Make a service appointment.

3) Analysis report release and view: Once the authorized BCI service providers obtain the user's neurophysiological data, the data will be analyzed to generate a mental state assessment of the corresponding user. As per Figure 11, the analysis report can be updated to the specified account address. Once the reports are created or updated, the system will generate a record ID and hash of the record and save them in the blockchain (Figure 12). The analysis reports are encrypted with a symmetric key. For the project managers, they can view the reports of their subordinate workers according to the worker's public key.

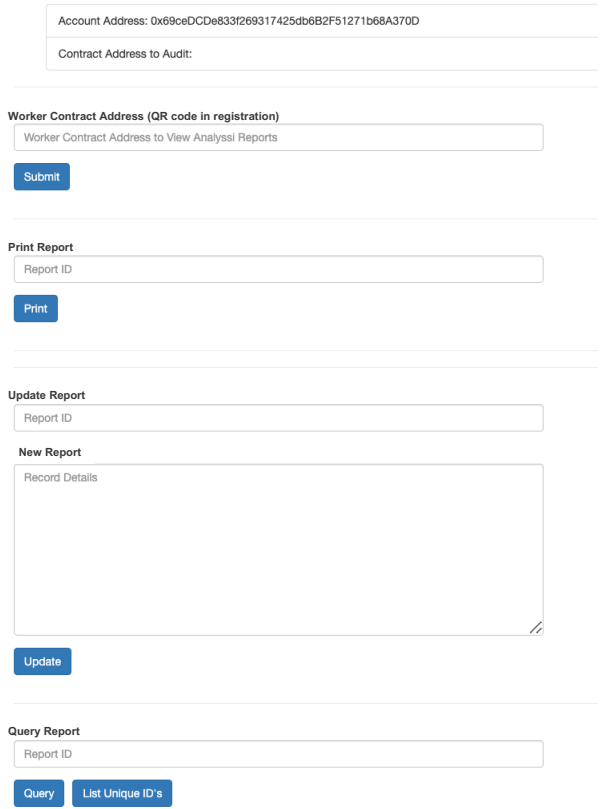

**FIGURE 11.** Update the analysis report.

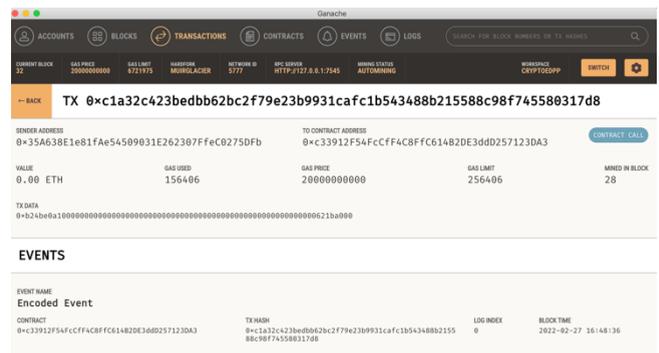

**FIGURE 12.** Compiled details in blockchain.

The proposed protocol is a blockchain-based system in which workers completely own their personal neurophysiological data, and that data storage and sharing among various stakeholders in the system is secure and reliable. Besides worker ownership, data storage security is another important characteristic of the proposed blockchain. The data is encrypted using hash algorithms and the worker's basic information filled in when registering is used to encrypt his or her public key. These data are stored in cloud storage secured through the blockchain. The information of personal neurophysiological data, such as the storage address, the hash value, and access permissions, are recorded onto blocks. These blocks are generated on the blockchain over time. Each block contains a hash of the preceding block, and any modification to the original data will change its hash value, which protects the data from being tampered with once they are written into the blockchain. The decentralized characteristic of blockchain guarantees the data to be updated in real-time since every network node participates in the data storage. These mechanisms ensure that the data are securely collected and permanently stored on a cloud server.

Moreover, the proposed blockchain system ensures secure data storing and sharing and provides great opportunities in the BCI applications business. The construction workers can choose to share their neurophysiological data into the public research domain anonymously and will be rewarded with a certain number of tokens. Blockchain technology helps workers and BCI service providers quickly and securely authenticate permissions for free data access and sharing, thus reducing the cost of data transmissions. With opportunities becoming available for a wide range of BCI service developers rather than being concentrated in the hands of a few institutions capable of collecting brain signals, this will level the playing field and create the opportunities for a new generation of entrepreneurs to build BCI service cases. Meantime, the users will receive better service.

## VI. CONCLUSION

With the tremendous growth of BCI applications, numerous private neurophysiological data are collected and stored in a centralized way, which poses security and privacy risks from cyber-attacks. Therefore, this study proposes a storage scheme and service framework based on deep learning and blockchain technologies for storing and sharing private neurophysiological data in a secure and privacy-preserving manner. To accomplish this goal, firstly, we utilized GAN to produce synthetic brain signals and applied the paradigm of transfer learning to classify EEG signals accurately. The proposed method to generate augmented EEG signals not only provides sufficient privacy-preservation safeguarding data leakage by concealing the sensitive patterns of the original data, but also gives superior classification performance revealed by a case study. Sensitivity analysis of the ratio of artificial to real data and GAN training epoch showed that data augmented with a larger size of low-quality artificially generated samples could achieve a favorable trade-off between data quality and data privacy. Next, we introduced a privacy-aware protocol that ensures valuable neurophysiological data can flow safely and conveniently in a blockchain-enabled system. We described the application scenarios with an Ethereum blockchain prototype running smart contracts and analyzed its advantages, particularly in terms of secure data storage. In our scheme, the primary purpose of storing and sharing data in the blockchain is to ensure user ownership, tamper-proof, privacy protection and convenient interoperability. Furthermore, the blockchain-based system provides lots of potential in the BCI applications industry. The sharing of data will further help construction workers become active BCI users [46], improve service quality[47] and give better decision support to service providers[48].

The proposed protocol is not only limited to the construction industry but also can be more widely applied in the occupational environment. Future work in this area includes exploring more scenarios including various stakeholders, such as insurance companies and government departments, etc., and getting properly implementable blockchain apps in the future which will contain all details of stakeholders with the respective condition for their smart contacts. Also, surveys are encouraged to be done to investigate the views of the proposed blockchain-based protocol from the perspective of actual construction workers and practitioners.

## REFERENCES

1. Aryal, A., A. Ghahramani, and B. Becerik-Gerber, Monitoring fatigue in construction workers using physiological measurements. Automation in Construction, 2017. 82: p. 154-165.
2. Wang, D., et al., Monitoring workers' attention and vigilance in construction activities through a wireless and wearable electroencephalography system. Automation in Construction, 2017. 82: p. 122-137.
3. Chen, J., X. Song, and Z. Lin, Revealing the "Invisible Gorilla" in construction: Estimating construction safety through mental workload assessment. Automation in Construction, 2016. 63: p. 173-183.
4. Jebelli, H., S. Hwang, and S. Lee, EEG-based workers' stress recognition at construction sites. Automation in Construction, 2018. 93: p. 315-324.
5. Hwang, S., et al., Measuring Workers' Emotional State during Construction Tasks Using Wearable EEG. Journal of Construction Engineering and Management, 2018. 144.
6. Awolusi, I., et al., Enhancing Construction Safety Monitoring through the Application of Internet of Things and Wearable Sensing Devices: A Review. 2019. 530-538.
7. Silversmith, D., et al., Plug-and-play control of a brain-computer interface through neural map stabilization. Nature biotechnology, 2021. 39.
8. Lotte, F., Signal Processing Approaches to Minimize or Suppress Calibration Time in Oscillatory Activity-Based Brain–Computer Interfaces. Proceedings of the IEEE, 2015. 103: p. 871-890.
9. Samek, W., F. Meinecke, and K.-R. Müller, Transferring Subspaces Between Subjects in Brain--Computer Interfacing. IEEE Transactions on Biomedical Engineering, 2013. 60: p. 2289-2298.
10. Debie, E. and N. Moustafa, A Privacy-Preserving Generative Adversarial Network Method for Securing EEG Brain Signals. 2020.
11. Lotte, F., C. Guan, and K. Ang, Comparison of Designs Towards a Subject-Independent Brain-Computer Interface based on Motor Imagery. Vol. 2009. 2009. 4543-6.
12. Zheng, X., et al., Blockchain-based Personal Health Data Sharing System Using Cloud Storage. 2018. 1-6.

13. Denning, T., Y. Matsuoka, and T. Kohno, Neurosecurity: Security and privacy for neural devices. Neurosurgical focus, 2009. 27: p. E7.
14. Zhang, Q. and Y. Liu, Improving brain computer interface performance by data augmentation with conditional Deep Convolutional Generative Adversarial Networks. 2018.
15. Lin, C., et al., A New Transitively Closed Undirected Graph Authentication Scheme for Blockchain-Based Identity Management Systems. IEEE Access, 2018. PP: p. 1-1.
16. Mamoshina, P., et al., Converging blockchain and next-generation artificial intelligence technologies to decentralize and accelerate biomedical research and healthcare. Oncotarget, 2018. 9: p. 5665-5690.
17. Lin, C., et al., BSeIn: A blockchain-based secure mutual authentication with fine-grained access control system for industry 4.0. Journal of Network and Computer Applications, 2018. 116: p. 42-52.
18. Zyskind, G., et al., Decentralizing Privacy: Using Blockchain to Protect Personal Data. 2015. 180-184.
19. Shahnaz, A., U. Qamar, and D. Khalid, Using Blockchain for Electronic Health Records. IEEE Access, 2019. PP: p. 1-1.
20. Azaria, A., et al., MedRec: Using Blockchain for Medical Data Access and Permission Management. 2016. 25-30.
21. Griggs, K., et al., Healthcare Blockchain System Using Smart Contracts for Secure Automated Remote Patient Monitoring. Journal of Medical Systems, 2018. 42.
22. Kuo, T.-T. and L. Ohno-Machado, ModelChain: Decentralized Privacy-Preserving Healthcare Predictive Modeling Framework on Private Blockchain Networks. 2018.
23. Yue, X., et al., Healthcare Data Gateways: Found Healthcare Intelligence on Blockchain with Novel Privacy Risk Control. Journal of medical systems, 2016. 40: p. 218.
24. Ekblaw, A., et al., A Case Study for Blockchain in Healthcare: "MedRec" prototype for electronic health records and medical research data.
25. K, A., A Blockchain-Based Smart Contract System for Healthcare Management. Electronics, 2020. 9: p. 94.
26. R, M. and B. G, Securing Sensitive Data in Body Area Sensor Network Using Blockchain Technique. 2020. 1-5.
27. Bak, S., Y. Pyo, and J. Jeong, Protection of EEG Data using Blockchain Platform. 2019. 1-3.
28. Sun, J., et al., A blockchain-based framework for electronic medical records sharing with fine-grained access control. PloS one, 2020. 15: p. e0239946.
29. Chen, Y., et al., Blockchain-Based Medical Records Secure Storage and Medical Service Framework. Journal of Medical Systems, 2018. 43.
30. Li, Y., et al., A Security Transmission and Storage Solution about Sensing Image for Blockchain in the Internet of Things. Sensors, 2020. 20: p. 916.
31. Zhou, X., et al., Hazard differentiation embedded in the brain: A near-infrared spectroscopy-based study. Automation in Construction, 2021. 122(C): p. 103473.
32. Schirrmeister, R., et al., Deep learning with convolutional neural networks for EEG decoding and visualization: Convolutional Neural Networks in EEG Analysis. Human Brain Mapping, 2017. 38.
33. Craik, A., Y. He, and J. Contreras-Vidal, Deep learning for Electroencephalogram (EEG) classification tasks: A review. Journal of Neural Engineering, 2019. 16.
34. Singh, R., et al., SeizSClas: An Efficient and Secure Internet of Things Based EEG Classifier. IEEE Internet of Things Journal, 2020. PP: p. 1-1.
35. Li, X., et al., Convolutional neural networks based transfer learning for diabetic retinopathy fundus image classification. 2017. 1-11.
36. Goodfellow, I., et al., Generative Adversarial Nets. ArXiv, 2014.
37. Hartmann, K., R. Schirrmeister, and T. Ball, EEG-GAN: Generative adversarial networks for electroencephalograhic (EEG) brain signals. 2018.
38. Lotte, F., Generating Artificial EEG Signals To Reduce BCI Calibration Time. 5th International Brain-Computer Interface Workshop, 2011.
39. Choi, E., et al., Generating Multi-label Discrete Patient Records using Generative Adversarial Networks. 2017.
40. Esteban, C., S. Hyland, and G. Rätsch, Real-valued (Medical) Time Series Generation with Recurrent Conditional GANs. 2017.
41. Christidis, K. and M. Devetsikiotis, Blockchains and Smart Contracts for the Internet of Things. IEEE Access, 2016. 4: p. 1-1.
42. Tallon-Baudry, C., et al., Stimulus Specificity of Phase-Locked and Non-Phase-Locked 40 Hz Visual Responses in Human. The Journal of neuroscience : the official journal of the Society for Neuroscience, 1996. 16: p. 4240-9.
43. Oostenveld, R., et al., FieldTrip: open source software for advanced analysis of MEG, EEG, and invasive electrophysiological data. Computational Intelligence Neuroscience Letters, 2011. 2011: p. 156869-156869.
44. Zhou, X., P.-C. Liao, and Q. Xu, Reinvestigation of the Psychological Mechanisms of Construction Experience on Hazard Recognition Performance. Human factors, 2022: p. 187208211066666-187208211066666.
45. Cohen, M. and R. Gulbinaite, Five methodological challenges in cognitive electrophysiology. NeuroImage, 2013. 85.
46. Huba, N. and Y. Zhang, Designing Patient-Centered Personal Health Records (PHRs): Health Care Professionals' Perspective on Patient-Generated Data. Journal of Medical Systems, 2012. 36(6): p. 3893-3905.
47. Xia, C., et al., Risk Analysis and Enhancement of Cooperation Yielded by the Individual Reputation in the Spatial Public Goods Game. IEEE Systems Journal, 2016. PP: p. 1-10.
48. Simpao, A., et al., A Review of Analytics and Clinical Informatics in Health Care. Journal of medical systems, 2014. 38: p. 45.